\documentclass[journal ]{new-aiaa}
\usepackage[utf8]{inputenc}
\usepackage{textcomp}

\usepackage{graphicx}
\usepackage{amsmath}
\usepackage[version=4]{mhchem}
\usepackage{siunitx}
\usepackage{longtable,tabularx}
\usepackage{bm}
\title{Energy Transition Domain and Its Application in Constructing Gravity-Assist Escape Trajectories}

\author{Shuyue Fu \footnote{PhD Candidate, Shen Yuan Honors College, School of Astronautics, fushuyue@buaa.edu.cn.}, Xiaowen Liu \footnote{PhD Candidate, School of Astronautics, Liuxiaowen2025@buaa.edu.cn.}, Di Wu \footnote{Associate Professor, School of Astronautics, wudi2025@buaa.edu.cn, Member AIAA.}, Peng Shi\footnote{Professor, School of Astronautics, shipeng@buaa.edu.cn.}, and Shengping Gong \footnote{Professor, School of Astronautics, gongsp@buaa.edu.cn, Senior Member AIAA (Corresponding Author).}}
\affil{Beihang University, Beijing, 100191, People's Republic of China}
\affil{State Key Laboratory of High-Efficiency Reusable Aerospace Transportation Technology, Beijing, 102206, People's Republic of China}
\affil{Key Laboratory of Spacecraft Design Optimization $\&$ Dynamic Simulation Technologies, Ministry of Education, Beijing, 100191, People's Republic of China}

\begin{document}

\maketitle

\begin{abstract}
This Note proposes the concept and theory of energy transition domain (ETD) defined by the mechanical energy of spacecraft in the Earth-Moon planar circular restricted three-body problem (PCR3BP) inspired by the pioneering work from Ano{\`e} et al. (2024) on the ETD defined by the two-body energy with respect to the secordary body in the PCR3BP. An effective construction method of gravity-assist escape trajectories is then proposed. Firstly, the concept of the ETD defined by the mechanical energy is presented, and its dependency on the Jacobi energy is analyzed. This dependency may provide prior knowledge about selecting the range of the Jacobi energy in the construction of escape trajectories. Then, gravity-assist escape trajectories departing from the 167 km low Earth orbit and 36000 km geosynchronous Earth orbit are constructed based on the ETD. The initial states are selected in the sphere of influence of the Moon, and the trajectories are searched from the forward and backward integration. Finally, the obtained solutions are presented and analyzed.
\end{abstract}

\section{Introduction}
\lettrine{E}{scape} and capture dynamics is a fundamental problem in celestial mechanics and astrodynamics \cite{todorovic2020arches,li2019energy,luo2014constructing,FU2025}. In particular, studies on escape dynamics provide an explanation of asteroid (space debris) evolution \cite{todorovic2020arches,jiao2024asteroid} and provide prior knowledge to aid in constructing escape trajectories that support interplanetary and asteroid exploration \cite{wang2017transfer,mingotti2011earth}. For the engineering application, a key challenge is to construct escape trajectories that satisfy mission requirements. Firstly, the model to construct escape trajectories and the criteria to identify them should be selected. For the restricted two-body problem \cite{Battin1999}, escape trajectories are equivalent to parabolic or hyperbolic trajectories with respect to the central body with a constant of two-body energy due to the existence of the closed-form solutions. Furthermore, the introduction of multi-body effects (i.e., adopting the restricted three-body problem \cite{zotos2015classifying,FU2025} or restricted four-body problem \cite{qi2016study,casalino2020design}) offers greater potential for generating new families of escape trajectories, e.g., gravity-assist escape trajectories \cite{pasquale2024systematic}. To further explore escape dynamics in the multi-body problems, we adopt one of the restricted three-body problems, namely, the Earth-Moon planar restricted three-body problem (PCR3BP), as the dynamical model. In the Earth-Moon PCR3BP, however, there are no closed-form solutions existing \cite{fu2025analytical}. Therefore, the definition and identification of escape trajectories should be adjusted. For the Earth-Moon PCR3BP, escape trajectories can be defined as trajectories escaping from infinity with respect to the Earth-Moon barycenter when the integration time increases to infinity \cite{zotos2015classifying,FU2025}. Because there are no closed-form solutions existing, numerical criterion to identify this type of trajectory should be developed. Zotos \cite{zotos2015classifying} adopted a distance criterion to identify escape trajectories. When the trajectories intersect the disk of the PCR3BP system (i.e., the distance between the test body (spacecraft) and the Earth-Moon barycenter exceeds a threshold), the trajectories can be considered as escape trajectories. This criterion has been used widely in the identification of natural and artificial escape trajectories \cite{zotos2015classifying,fu2025four}. However, this criterion only use the geometric properties of escape trajectories, lacking a use of their dynamical properties. Based on this criterion, Fu and Gong \cite{FU2025} introduced an energy criterion. They used the concept of the mechanical energy of spacecraft, proposed by Broucke \cite{broucke1968periodic} and Qi and Xu \cite{qi2015mechanical}, and established a link between escape trajectories and this mechanical energy. By combining the energy criterion (the value of mechanical energy is positive) with the aforementioned distance criterion, they proposed a new escape criterion. As a result, in the test cases, the new escape criterion can further exclude trajectories which have not already escaped compared to the criterion adopted by Zotos \cite{zotos2015classifying}. Therefore, we adopt the criterion developed by Fu and Gong \cite{FU2025} to identify escape trajectories. Based on this criterion, we aim to extend the analysis and provide further insights into escape dynamics, which is the first purpose of this Note. 

Once the dynamical model to construct escape trajectories and criterion to identify escape trajectories are determined, we focus on the type of escape trajectories. Escape trajectories departing from several parking orbits have been studied, including the circular Earth orbits \cite{fu2025four,FU2025}, circular Moon orbits \cite{qian2016energy}, and three-body periodic orbits \cite{mccarthy2021leveraging,boudad2022departure}. In this Note, we focus on escape trajectories departing from circular Earth parking orbits (low Earth orbit (LEO) and geosynchronous Earth orbit (GEO)) because it is more realistic at the current stage to plan interplanetary or asteroid transfer missions from such orbits. For this type of escape trajectory in the Earth-Moon PCR3BP, the trajectories can be further categorized into direct escape trajectories and gravity-assist escape trajectories. In particular, we focus on the lunar gravity assist (LGA) as it generally causes a larger variation in the mechanical energy than the Earth gravity assist \cite{qi2015mechanical}. Similar to the effects of the LGA on the reduction of impulses of the Earth-Moon transfers \cite{qi2017optimal,oshima2019low,wang2025mechanism}, it is expected to generate low-energy solutions of escape trajectories using the LGA. However, also because the absence of the closed-form solutions in the Earth-Moon PCR3BP, it is difficult to obtain the initial states of gravity-assist trajectories (also gravity-assist escape trajectories) directly \cite{yang2025deep}. Moreover, the grid search method \cite{FU2025} generates escape trajectories including direct escape and gravity-assist escape. Therefore, this method can not effectively exclude the direct escape trajectories. Motivated by the aforementioned discussion, the second purpose of this Note is to develop an effective method to construct gravity-assist escape trajectories.

In the criterion developed by Fu and Gong \cite{FU2025}, there is a key indicator to identify escape trajectories, i.e., the mechanical energy of spacecraft. A positive value of the mechanical energy is one of the identification conditions. Inspired by the pioneering work on the energy transition domain (ETD) defined by the two-body energy with respect to the secondary body in the PCR3BP developed by Ano{\`e} et al. \cite{anoe2024ballistic}, we consider a critical case where the value of the mechanical energy of spacecraft is equal to zero, and propose a concept of ETD defined by the mechanical energy of spacecraft. Differing from their work on the ETD defined by the two-body energy with respect to the secondary body used to analyze ballistic capture, our proposed ETD is used to analyze escape dynamics in the Earth-Moon PCR3BP. The dependence of the ETD on the Jacobi energy is analyzed, and an interesting phenomenon is identified, which may provide prior knowledge to aid in constructing escape trajectories. Then, a construction method of gravity-assist escape trajectories based on the ETD is proposed. Simulation results verify the effectiveness of the proposed method, confirming that the LGA can help reduce the impulse of escape trajectories compared to direct escape.

The rest of this Note is organized as follows. Section \ref{sec2} presents the mathematical background of this work. Section \ref{sec3} introduces the concept of the ETD defined by the mechanical energy of spacecraft and its dependency on the Jacobi energy. Section \ref{sec4} proposes the method to construct gravity-assist escape trajectories based on the ETD. The effectiveness of the method is verified by the simulations. Finally, conclusions are drawn in Section \ref{sec5}.

\section{Mathematical Background}\label{sec2}
This section introduces the mathematical background of this work, including the Earth-Moon PCR3BP, the concepts of energy in the Earth-Moon PCR3BP, and the escape criterion to identify escape trajectories in the Earth-Moon PCR3BP.
\subsection{Earth-Moon PCR3BP}\label{subsec2.1}
In this Note, the Earth-Moon PCR3BP is adopted to investigate escape dynamics and construct gravity-assist escape trajectories. This model can provide a higher fidelity than the Earth- or Moon-centered two-body problem and the Earth-Moon patched two-body problem while maintaining lower complexity compared to the Sun-Earth/Moon restricted four-body problem and an ephemeris model \citep{mccarthy2021leveraging,FU2025}. The details about this model can be found in Ref. \cite{FU2025}. The dynamical equations are expressed as:
\begin{equation}
\left[ {\begin{array}{*{20}{c}}
{\begin{array}{*{20}{c}}
{\dot x}\\
{\dot y}
\end{array}}\\
{\begin{array}{*{20}{c}}
{\dot u}\\
{\dot v}
\end{array}}
\end{array}} \right] = \left[ {\begin{array}{*{20}{c}}
{\begin{array}{*{20}{c}}
u\\
v
\end{array}}\\
{\begin{array}{*{20}{c}}
{2v + \frac{{\partial {\Omega _3}}}{{\partial x}}}\\
{ - 2u + \frac{{\partial {\Omega _3}}}{{\partial y}}}
\end{array}}
\end{array}} \right]\label{eq1}
\end{equation}

\begin{equation}
{\Omega _3} = \frac{1}{2}\left[ {{x^2} + {y^2} + \mu \left( {1 - \mu } \right)} \right] + \frac{{1 - \mu }}{{{r_1}}} + \frac{\mu }{{{r_2}}}\label{eq2}
\end{equation}
where $\bm{X} = \left[ x, y, u, v\right]^{\text{T}}$ denotes the orbital state, and ${\Omega _3}$ is the effective potential of the Earth-Moon PCR3BP. The parameter $\mu $ denotes the Earth-Moon mass parameter. The distance between the spacecraft and the Earth ($r_1$), the Moon ($r_2$), and the Earth-Moon barycenter ($r$) are expressed as:
\begin{equation}
{r_1} = \sqrt {{{\left( {x + \mu } \right)}^2} + {y^2}} \text{ }\text{ }\text{ }\text{ }{r_2} = \sqrt {{{\left( {x + \mu - 1} \right)}^2} + {y^2}}\text{ }\text{ }\text{ }\text{ }r=\sqrt{x^2+y^2}\label{eq3}
\end{equation}
For numerical integration of trajectories in the Earth-Moon PCR3BP, we use the variable step-size, variable order (VSVO) Adams-Bashforth-Moulton algorithm with absolute and relative tolerances set to $1 \times 10^{-13}$, performed via MATLAB®'s ode113 command \cite{oshima2021capture}. The specific values of the parameters used in this Note can be found in Ref. \cite{FU2025}. Subsequently, the concepts of energy in the Earth-Moon PCR3BP is introduced.

\subsection{Energy in the Earth-Moon PCR3BP}\label{subsec2.2}
This subsection introduces two concepts of energy in the Earth-Moon PCR3BP, namely, the Jacobi energy and the mechanical energy of spacecraft. Along the trajectories in the Earth-Moon PCR3BP, there exists a constant denoted as the Jacobi energy, which is expressed as:
\begin{equation}
C = -\left( {{u^2} + {v^2}} \right) +  \left( {{x^2} + {y^2}} \right) + \frac{{2(1 - \mu) }}{{{r_1}}} + \frac{2\mu }{{{r_2}}} + \mu \left( {1 - \mu } \right) \label{eq4}
\end{equation}

\begin{figure}[h]
\centering
\includegraphics[width=0.45\textwidth]{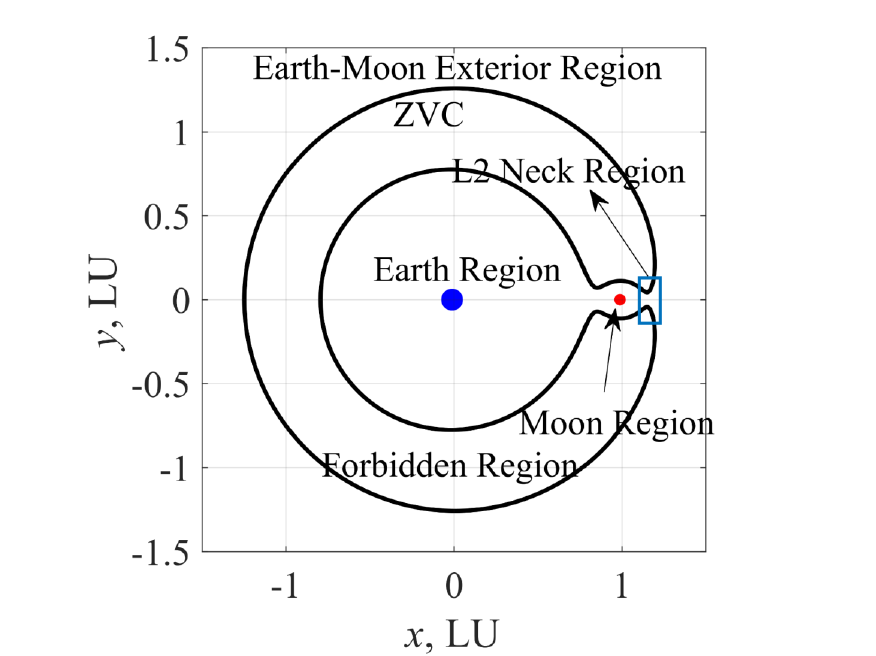}
\caption{Reachable region under $C_0<C_{\text{L2}}$.}
\label{neck_region}
\end{figure}

Using the concept of the Jacobi energy, the Hill region can be defined as:
\begin{equation}
H = \left\{ {\left( {x,{\text{ }}y} \right)|\left( {{x^2} + {y^2}} \right) + \frac{{2(1 - \mu) }}{{{r_1}}} + \frac{2\mu }{{{r_2}}} + \mu \left( {1 - \mu } \right) \geq {C}} \right\} \label{eq5}
\end{equation}
here $C$ denotes a given value of the Jacobi energy. Equation \eqref{eq5} defines the reachable region of the spacecraft under the given value of the Jacobi energy. Furthermore, the zero-velocity curve (ZVC) under the given value of the Jacobi energy is defined as:

\begin{equation}
\text{ZVC} = \left\{ {\left( {x,{\text{ }}y} \right)|\left( {{x^2} + {y^2}} \right) + \frac{{2(1 - \mu) }}{{{r_1}}} + \frac{2\mu }{{{r_2}}} + \mu \left( {1 - \mu } \right) = {C}} \right\} \label{eq6}
\end{equation}
When $C$ is less than the Jacobi energy of the L2 libration point ($C_{\text{L2}}=3.184158216376\text{ } {\left( {{\text{LU/TU}}} \right)^2}$), the neck region near the L2 libration point opens and the spacecraft is able to transfer from the Earth region to the Moon region and the Earth-Moon exterior region, as shown in Fig. \ref{neck_region}. Therefore, to construct escape trajectories from the Earth region, the necessary condition is that the Jacobi energy of the trajectories is less than $C_{\text{L2}}$ \cite{villac2003escaping}.

However, the condition that $C<C_{\text{L2}}$ is not sufficiently accurate to identify escape trajectories \cite{FU2025}. To further identify escape trajectories in the Earth-Moon PCR3BP, the concept the mechanical energy of spacecraft ($E$) is presented. The mechanical energy of spacecraft can be expressed as \cite{qi2015mechanical}:
\begin{equation}
E = \frac{1}{2}\left( {{x^2} + {y^2} + {u^2} + {v^2}} \right) + xv - uy - \frac{{1 - \mu }}{{{r_1}}} - \frac{\mu }{{{r_2}}} \label{eq7}
\end{equation}
This concept was firstly adopted by Broucke \cite{broucke1968periodic} in the search of periodic orbits, implying that when $E>0$, trajectories may escape. Using this concept, Qi and Xu \cite{qi2015mechanical} further investigated the theory of LGA in the PCR3BP. Although the aforementioned works \cite{broucke1968periodic,qi2015mechanical} suggested that $E>0$ corresponded to so-called “hyperbolic (or open) trajectory”, the theoretical foundation of using this criterion ($E>0$) to identify escape trajectories remained relatively weak. Based on their works, Fu and Gong \cite{FU2025} further established the escape criterion using this concept and approximate relationship between $E$ and two-body energy with respect to the Earth-Moon barycenter (detailed in Appendix A). Subsequently, the escape criterion is presented. 

\subsection{Escape Criterion}\label{subsec2.3}
Several definitions of escape have been proposed in the previous works \cite{topputo2008resonant,mingotti2011earth,oshima2021capture,qian2016energy,zotos2015classifying,FU2025,luo2020role}. Escape trajectories have been identified based on the two-body energy with respect to the Earth \cite{topputo2008resonant,mingotti2011earth}, the hyperbolic excess velocities with respect to the Earth \cite{oshima2021capture}, the distance from the Earth-Moon barycenter \cite{zotos2015classifying,fu2025four}, the concept of transit orbits \cite{qian2016energy}, and the combination of energy and distance \cite{FU2025,luo2020role}. In this Note, following the work of Ref. \cite{FU2025}, we define escape trajectories in the Earth-Moon PCR3BP as trajectories escaping to infinity relative to the Earth-Moon barycenter when the orbital integration time tends to infinity (i.e., $\text{d}r/\text{d}t>0$ for all times $t > t_1$, for some finite time $t_1$). Notably, this definition is suitable for the Earth-Moon PCR3BP. When the solar gravity perturbation is considered \cite{casalino2020design,FU20254993}, this definition should be further adjusted. To identify escape trajectories following the aforementioned definition, we adopt the escape criterion developed by Fu and Gong \cite{FU2025} to identify escape trajectories. This criterion is presented as follows:
\begin{enumerate}[label=(\roman*)]
\item The distance between the spacecraft and the Earth-Moon barycenter $r$ exceeds a given threshold, i.e., $r > R_d$.
\item The time derivative of $r$ satisfies $  \text{d}r/\text{d}t > 0$.
\item The mechanical energy of spacecraft satisfies $ E > 0$.
\end{enumerate}
When the states of trajectories satisfy these three conditions simultaneously, the trajectories can be considered as escape trajectories (escape has already taken place). This criterion is similar to the criterion proposed by Luo \cite{luo2020role}. However, this criterion uses the concept of $E$ while Luo's criterion used the two-body energy with respect to the primary body (in the Earth-Moon PCR3BP, the primary body refers to the Earth). For identifying escape trajectories from the Earth-Moon systems, the physical meaning of $E$ is more intuitive as it considers both the effects of the Earth and the Moon. In this Note, we still set $R_d$ to $10\text{ LU}$. When escape trajectories in the Earth-Moon PCR3BP are identified, the trajectory segment in the region where the motion of the spacecraft is mainly dominated by the gravity of the Earth and the Moon \cite{qi2016study} is expected to be continued to higher-fidelity models. 

From the escape criterion developed by Fu and Gong \cite{FU2025}, it can be observed that the energy condition ($E>0$) is important to identify escape trajectories. Therefore, we focus on the critical case, i.e., $E=0$, and propose the concept of the ETD defined by $E$ following the idea of Ano{\`e} et al. \cite{anoe2024ballistic} on the ETD defined by the two-body energy with respect to the secondary body.

\section{Energy Transition Domain Defined by $E$}\label{sec3}
This section introduces the concept of ETD defined by $E$ and the corresponding dependency with respect to $C$.
\subsection{Concept of ETD}\label{subsec3.1}
The ETD defined by $E$ describes a set of $\left(x,\text{ }y\right)$ that can satisfy $E=0$ under a given value of $C$. Then, we provide a rigorous mathematical definition of ETD defined by $E$. Using the following parameterization of the orbital state:
\begin{equation}
\left\{ \begin{gathered}
  x = r\cos \alpha {\text{ }\text{ }\text{ }}y = r\sin \alpha {\text{   }\text{ }\text{ }}u = V\sin \sigma {\text{   }\text{ }\text{ }}v = V\cos \sigma  \hfill \\
  V = \sqrt { - C + {r^2} + \frac{{2\left( {1 - \mu } \right)}}{{{r_1}}} + \frac{{2\mu }}{{{r_2}}} + \mu \left( {1 - \mu } \right)}  \hfill \\ 
\end{gathered}  \right. \label{eq8}
\end{equation}
the $E$ can be expressed as ($\alpha$ and $\sigma$ denotes two parameters satisfying $\alpha \in \left[0,\text{ }2\pi\right)$ and $\sigma \in \left[0,\text{ }2\pi\right)$):
\begin{equation}
E = \frac{1}{2}\left( {{r^2} + {V^2} + 2rV\cos \left( {\alpha  + \sigma } \right)} \right) - \frac{{1 - \mu }}{{{r_1}}} - \frac{\mu }{{{r_2}}} \label{eq9}
\end{equation}
Because $-1\leq\cos \left( {\alpha  + \sigma } \right)\leq 1$, $E$ satisfies:
\begin{equation}
 E_{\text{lower}} \leq E\leq E_{\text{upper}} \label{eq10}
\end{equation}
where:
\begin{equation}
 {E_{{\text{lower}}}} = \frac{1}{2}{\left( {V - r} \right)^2} - \frac{{1 - \mu }}{{{r_1}}} - \frac{\mu }{{{r_2}}}\label{eq11}
\end{equation}
\begin{equation}
 {E_{{\text{upper}}}} = \frac{1}{2}{\left( {V + r} \right)^2} - \frac{{1 - \mu }}{{{r_1}}} - \frac{\mu }{{{r_2}}}\label{eq12}
\end{equation}
When satisfying:
\begin{equation}
{E_{{\text{lower}}}} \cdot {E_{{\text{upper}}}} \leq 0\label{eq13}
\end{equation}
we denote the pair $\left(x,\text{ }y\right)$ belongs to the ETD defined by $E$ under the given value of $C$ according to root existence theorem. Then, the expression of ETD defined by $E$ under the given value of $C$ can be presented as:
\begin{equation}
{\text{ETD}} = \left\{ {\left( {x,{\text{ }}y} \right)|{E_{{\text{lower}}}} \cdot {E_{{\text{upper}}}} \leq 0} \right\}\label{eq14}
\end{equation}

\subsection{Dependency of ETD defined by $E$ on $C$}\label{subsec3.2}
\begin{figure}[h]
\centering
\includegraphics[width=0.95\textwidth]{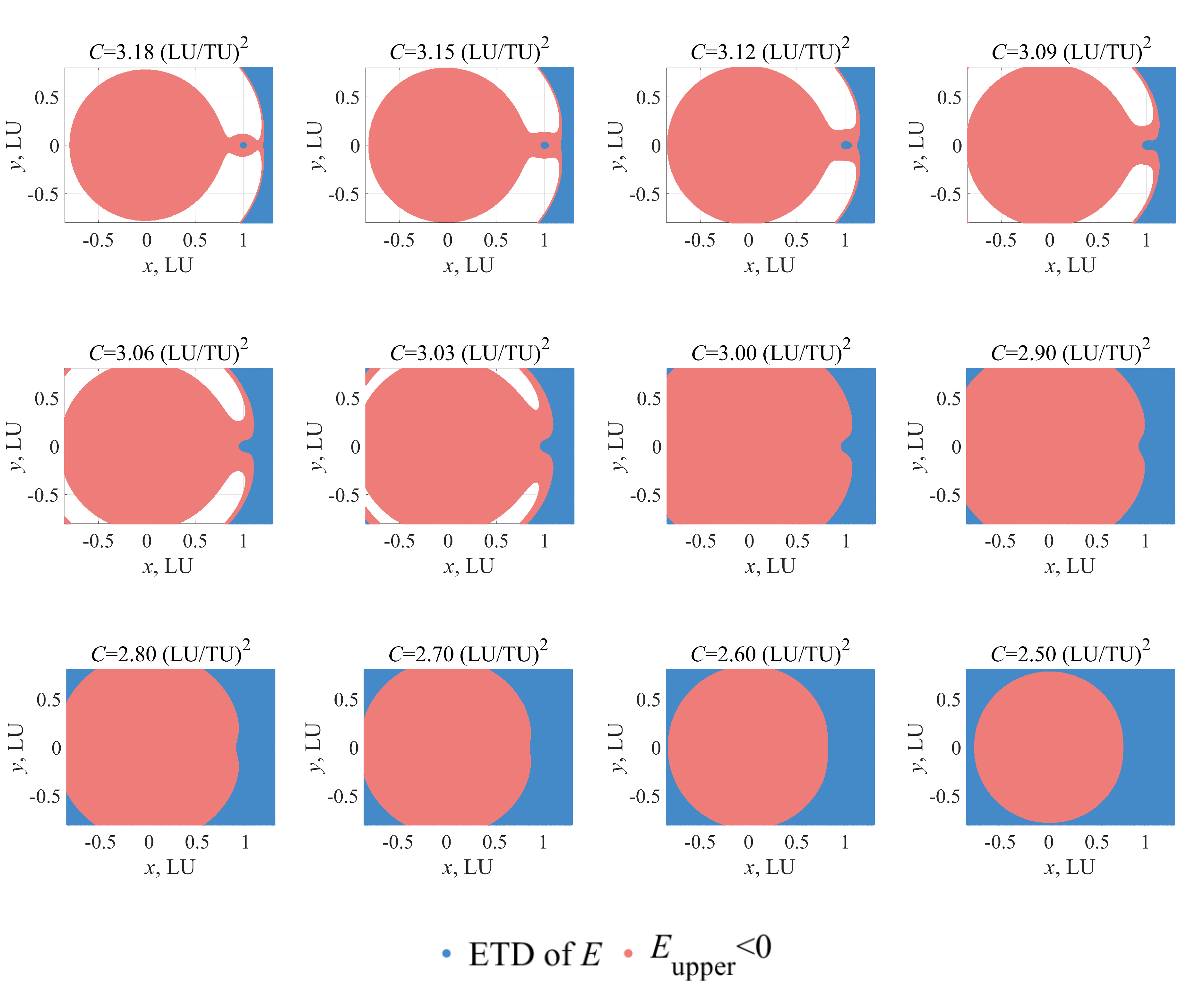}
\caption{Configurations of ETD with respect to $C$.}
\label{fig_con_C}
\end{figure}

Following the aforementioned discussion, the configuration of ETD defined by $E$ depends on the value of $C$. Here, we present the configurations with respect to $C$ in Fig. \ref{fig_con_C}. We select the range of $C$ as $C\in \left[2.50,\text{ }3.18\right] {\left( {{\text{LU/TU}}} \right)^2}$, because the necessary condition of escaping from the Earth region is $C<C_{\text{L2}}$ (i.e., the L2 region opens) and we aim to investigate escape trajectories with relatively low energy.

\begin{figure}[h]
  \centering
  \begin{minipage}{0.45\textwidth}
    \centering
    \includegraphics[width=0.55\textwidth]{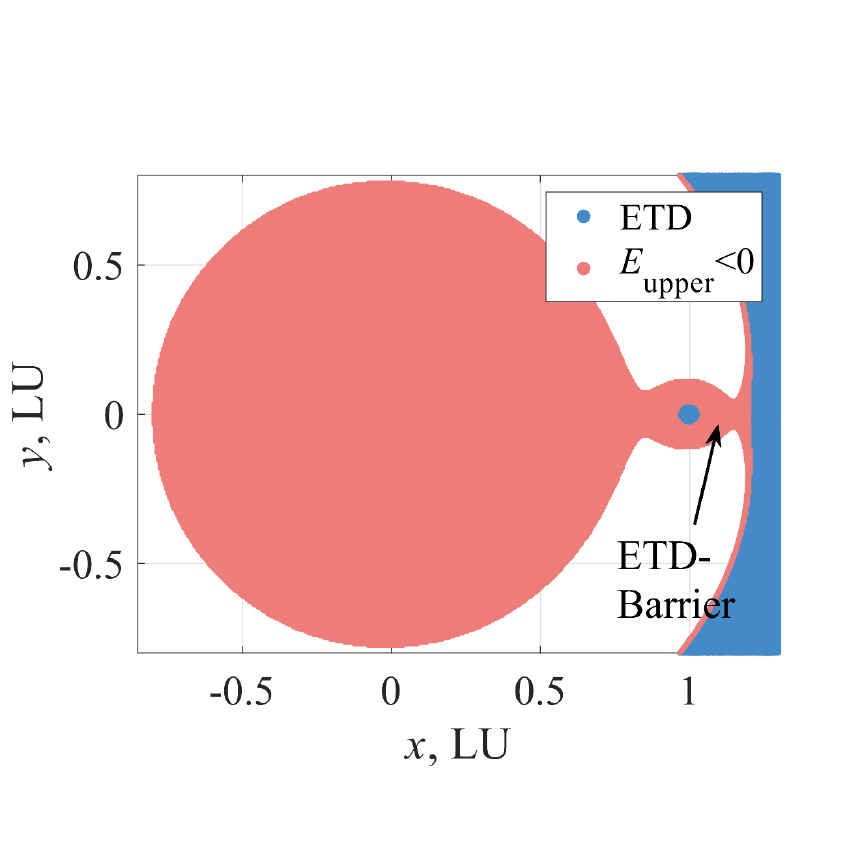}
    \caption{Barrier between two regions of ETD.}
    \label{fig_barrier}
  \end{minipage}%
  \hfill
  \begin{minipage}{0.45\textwidth}
    \centering
    \includegraphics[width=0.7\textwidth]{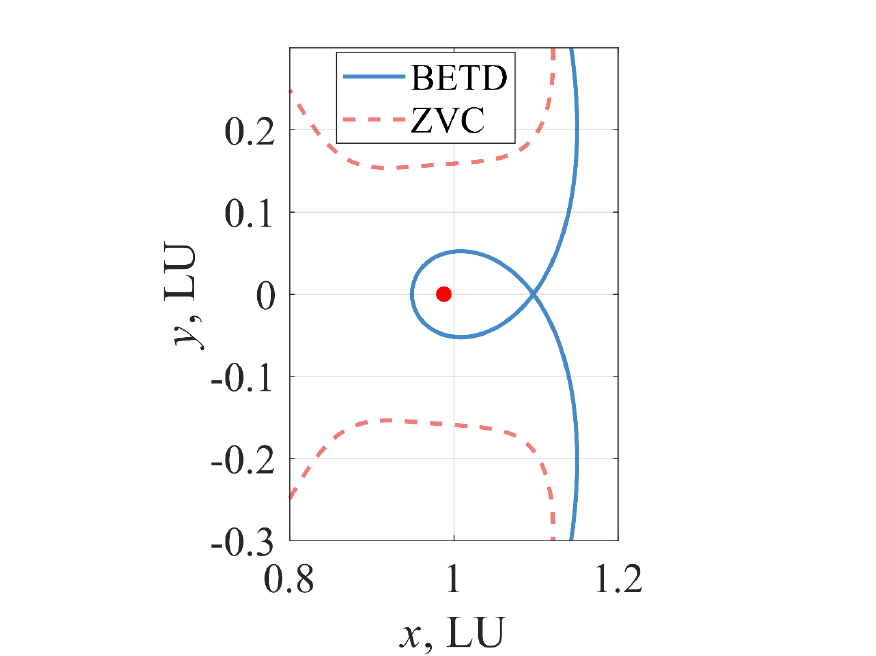}
    \caption{“Bifurcation point” of ETD.}
    \label{fig_bifurcation}
  \end{minipage}%
\end{figure}

In Fig. \ref{fig_con_C}, the blank region denotes the forbidden region, the blue region denotes the ETD, and the orange region denotes region corresponding to $E_\text{upper}<0$. It is observed that these three types of regions are symmetric about $y=0$. Notably, within the investigated range of $C$, no region corresponding to $E_\text{lower}>0$ exists. In contrast, the region corresponding to $E_\text{upper}<0$ always exists in the Earth region, while the ETD is mainly distributed in the Moon region and the Earth-Moon exterior region. Notably, there is also the ETD existing inside the Earth within the investigated range of $C$. However, we focus on the region distributed in the Moon region and the Earth-Moon exterior region because the region inside the Earth has no physical meaning. When the value of $C$ is relatively high (3.12-3.18), it is observed that the ETD is mainly divided into two regions, with a “barrier” existing between them, as shown in Fig. \ref{fig_barrier}. This is an interesting finding. It may imply that even under $C<C_\text{L2}$, trajectories may not effectively escape from the Earth region. Although the $E$ of the trajectory can satisfy $E=0$ after a LGA (in the ETD region in the Moon region), it becomes negative again when departing the the ETD region. In this case, the LGA may not be able to cause the variation in $E$ (i.e., $E<0 \to E>0$) effectively. As the value of $C$ decreases, we can find that two regions of ETD merge into one region. With further decrease, the region corresponding to $E_\text{upper}<0$ shrinks and the ETD region expands. Following this trend, a “bifurcation point” appears when the two ETD regions firstly come into contact, similar to Hill regions varying with $C$. Subsequently, we present detailed information about this critical case.

Following the idea adopted by Liu and Gong \cite{liu2017hill} of calculating the “bifurcation point”, we present the conditions that the “bifurcation point” $\left(x,\text{ }y\right)$ combined with $C$ in this case should satisfy:
\begin{equation}
\left\{ \begin{gathered}
  \frac{{\partial {E_{{\text{upper}}}}}}{{\partial x}} = 0 \hfill \\
  \frac{{\partial {E_{{\text{upper}}}}}}{{\partial y}} = 0 \hfill \\
  {E_{{\text{upper}}}} = 0 \hfill \\ 
\end{gathered}  \right.
\label{eq15}
\end{equation}
where:
\begin{equation}
\frac{{\partial {E_{{\text{upper}}}}}}{{\partial x}} = \left( {r + V} \right)\left( {\frac{x}{r} + \frac{{x - \frac{{\left( {1 - \mu } \right)\left( {x + \mu } \right)}}{{{r_1}^3}} - \frac{{\mu \left( {x + \mu  - 1} \right)}}{{{r_2}^3}}}}{V}} \right) + \frac{{\left( {1 - \mu } \right)\left( {x + \mu } \right)}}{{{r_1}^3}} + \frac{{\mu \left( {x + \mu  - 1} \right)}}{{{r_2}^3}}
\label{eq16}
\end{equation}
\begin{equation}
\frac{{\partial {E_{{\text{upper}}}}}}{{\partial y}} = \left( {r + V} \right)\left( {\frac{y}{r} + \frac{{y - \frac{{\left( {1 - \mu } \right)y}}{{{r_1}^3}} - \frac{{\mu y}}{{{r_2}^3}}}}{V}} \right) + \frac{{\left( {1 - \mu } \right)y}}{{{r_1}^3}} + \frac{{\mu y}}{{{r_2}^3}}
\label{eq17}
\end{equation}
Equation \eqref{eq15} is a nonlinear system of equations with respect to $x$ and $C$ ($y=0$ due to the symmetry). To solve this equation, we use MATLAB®'s fsolve command with Levenberg-Marquardt method and tolerances set to $1\times 10^{-20}$. The initial guess is set as $x=1.1\text{ LU}$ and $C=3.12\text{ }{\left( {{\text{LU/TU}}} \right)^2}$. After solving Eq. \eqref{eq15}, the detailed information about the “bifurcation point” is presented as follows (the superscript “*” denotes the quantities corresponding to the “bifurcation point”):
\begin{equation}
\left\{ \begin{gathered}
  x^* = {\text{1}}{\text{.096746490685516 LU}} \hfill \\
  y^* = 0 \hfill \\
  C^* = {\text{3}}{\text{.117819838289537 }}{\left( {{\text{LU/TU}}} \right)^2} \hfill \\ 
\end{gathered}  \right.
\label{eq18}
\end{equation}
Figure \ref{fig_bifurcation} presents the case under $C=C^*$. The term “BETD” denotes the boundary of the ETD. This figure illustrates the effectiveness of values presented in Eq. \eqref{eq15}. Therefore, when $C>C^*$, the ETD is mainly divided into two regions, and when $C<C^*$, these two regions merge into one region.

Based on the aforementioned discussion, we find that within the investigated range of $C$, escape trajectories from the Earth region (particular departing from LEO and GEO) starts with a negative $E$ (because the region corresponding to $E_\text{upper}<0$ always exists in the Earth region). To escape, these trajectories should pass through the ETD to achieve a positive $E$. This motivates the construction of escape trajectories from the Earth region based on the ETD. In particular, we focus on gravity-assist escape trajectories. Then, we propose the construction method and present the results.
\section{Constructing Gravity-Assist Escape Trajectories Based on ETD}\label{sec4}
In this section, we propose the construction method of gravity-assist escape trajectories based on the ETD. Construction results obtained from this method are then presented and discussed.
\subsection{Constraints of Gravity-Assist Escape Trajectories}\label{subsec4.1}
We consider cases where the trajectory departs from a circular Earth parking orbit by an Earth injection impulse ($\Delta{v}_i$) and finally escape from the Earth-Moon system (the trajectories satisfying the escape criterion mentioned in Section \ref{subsec2.3}) after the LGA. We consider the case using the tangential impulse, and the constraints can be expressed as \cite{FU2025,fu2025four}: 
\begin{equation}
{\bm{\psi }_i} = \left[ {\begin{array}{*{20}{c}}
  {{{\left( {{x_i} + \mu } \right)}^2} + {y_i}^2 - {{\left( {{R_{\text{E}}} + {h_i}} \right)}^2}} \\ 
  {\left( {{x_i} + \mu } \right)\left( {{u_i} - {y_i}} \right) + {y_i}\left( {{v_i} + {x_i} + \mu } \right)} 
\end{array}} \right] = \mathbf{0} \label{eq19}
\end{equation}
where $R_\text{E}$ denotes the Earth radius and $h_i$ denotes the orbital altitude of the Earth parking orbit. The subscript “\textit{i}” denotes the quantities corresponding to the departure point of the trajectory. In this Note, we considering two cases: $h_{i1}=167\text{ km}$ (corresponding to LEO) and $h_{i2}=36000\text{ km}$ (corresponding to GEO). When escape trajectories are identified, the impulse can be calculated by:
\begin{equation}
\Delta {v_i} = \sqrt {{{\left( {{u_i} - {y_i}} \right)}^2} + {{\left( {{v_i} + {x_i} + \mu } \right)}^2}}  - \sqrt {\frac{{1 - \mu }}{{{R_{\text{E}}} + {h_i}} }}  \label{eq20}
\end{equation}
Subsequently, we propose the construction method based on the ETD.
\subsection{Construction Method Based on ETD}\label{subsec4.2}
The construction method is detailed in the following texts, including generating initial states based on the ETD, performing forward integration to identify escape trajectories, and performing backward integration to search the states satisfying Eq. \eqref{eq19}.
\subsubsection{Generating Initial States Based on ETD}\label{subsubsec4.2.1}
The first step of the proposed construction method is to generate initial states of escape trajectories based on the ETD. The selected initial states satisfy $E=0$. To construct gravity-assist escape trajectories, we select the initial states corresponding to the Moon region. Therefore, the construction parameters are set as:
\begin{equation}
\bm{y} = {\left[ {{r_2},{\text{ }}{\alpha _{\text{M}}}} \right]^{\text{T}}}  \label{eq21}
\end{equation}
where $\alpha _{\text{M}}$ denotes the lunar phase angle. With these parameters, $\left(x,\text{ }y\right)$ can be obtained as:
\begin{equation}
x = {r_2}\cos {\alpha _{\text{M}}} + 1 - \mu {\text{ }\text{ }\text{ }\text{ }}y = {r_2}\sin {\alpha _{\text{M}}}  \label{eq22}
\end{equation}
We set $r_2$ as $r_2 \in \left[R_\text{M},\text{ }R_\text{SOI}\right]$ (dimensionless units) with a step-size of $\left(R_\text{SOI}-R_\text{M}\right)/1000$ (dimensionless units), where $R_\text{M}$ denotes the Moon radius and $R_\text{SOI}$ denotes the radius of SOI of the Moon. In this Note, we set $R_\text{SOI}$ to 66243 km \cite{pan2022research}. The parameter $\alpha _{\text{M}}$ is set as $\alpha _{\text{M}} \in \left[0,\text{ }2\pi\right)$ with a step-size of $\pi/3600$. Based on the discussion in Section \ref{sec3}, when $C>C^*$, the ETD is mainly divided into two regions, implying that trajectories may not effectively escape from the Earth region (an illustrative example is provided in Appendix B). Therefore, we set $C<C^*$ as $C \in \left[2.6,\text{ }3.1\right]\text{ } {\left( {{\text{LU/TU}}} \right)^2}$ with a step-size of 0.1 ${\left( {{\text{LU/TU}}} \right)^2}$. Given the values of $r_2$, $\alpha_\text{M}$, and $C$, we first select the pair $\left(x,\text{ }y\right)$ in the ETD (i.e., $\left(x,\text{ }y\right)$ satisfying Eq. \eqref{eq13}). Then, we calculate $\left(u,\text{ }v\right)$ corresponding to $\left(x,\text{ }y\right)$ in the ETD. The initial states should satisfy $E=0$. According to Eq. \eqref{eq9}, we can obtain:
\begin{equation}
\cos \left( {\alpha  + \sigma } \right) = \frac{{1 - \mu }}{{{r_1}rV}} + \frac{\mu }{{{r_2}rV}} - \frac{{{r^2} + {V^2}}}{{2rV}}  \label{eq23}
\end{equation}
Therefore, $\alpha  + \sigma$ can be obtained as:
\begin{equation}
\alpha  + \sigma  = \arccos \left( {\frac{{1 - \mu }}{{{r_1}rV}} + \frac{\mu }{{{r_2}rV}} - \frac{{{r^2} + {V^2}}}{{2rV}}} \right)\text{ }\text{ }{\text{ or }}\text{ }\text{ }2\pi  - \arccos \left( {\frac{{1 - \mu }}{{{r_1}rV}} + \frac{\mu }{{{r_2}rV}} - \frac{{{r^2} + {V^2}}}{{2rV}}} \right)  \label{eq24}
\end{equation}
Because $\alpha$ can be calculated by $\alpha=\text{atan2}\left(y,\text{ }x\right)$, we can obtain the value of the parameter $\sigma$:
\begin{equation}
 \sigma  = \arccos \left( {\frac{{1 - \mu }}{{{r_1}rV}} + \frac{\mu }{{{r_2}rV}} - \frac{{{r^2} + {V^2}}}{{2rV}}} \right)-\alpha\text{ }\text{ }{\text{ or }}\text{ }\text{ }2\pi  - \arccos \left( {\frac{{1 - \mu }}{{{r_1}rV}} + \frac{\mu }{{{r_2}rV}} - \frac{{{r^2} + {V^2}}}{{2rV}}} \right)-\alpha  \label{eq25}
\end{equation}
Then, according to Eq. \eqref{eq8}, we can obtain the initial states to construct gravity-assist escape trajectories.
\subsubsection{Performing Forward Integration to Identify Escape Trajectories}\label{subsubsec4.2.2}
\begin{figure}[h]
\centering
\includegraphics[width=0.9\textwidth]{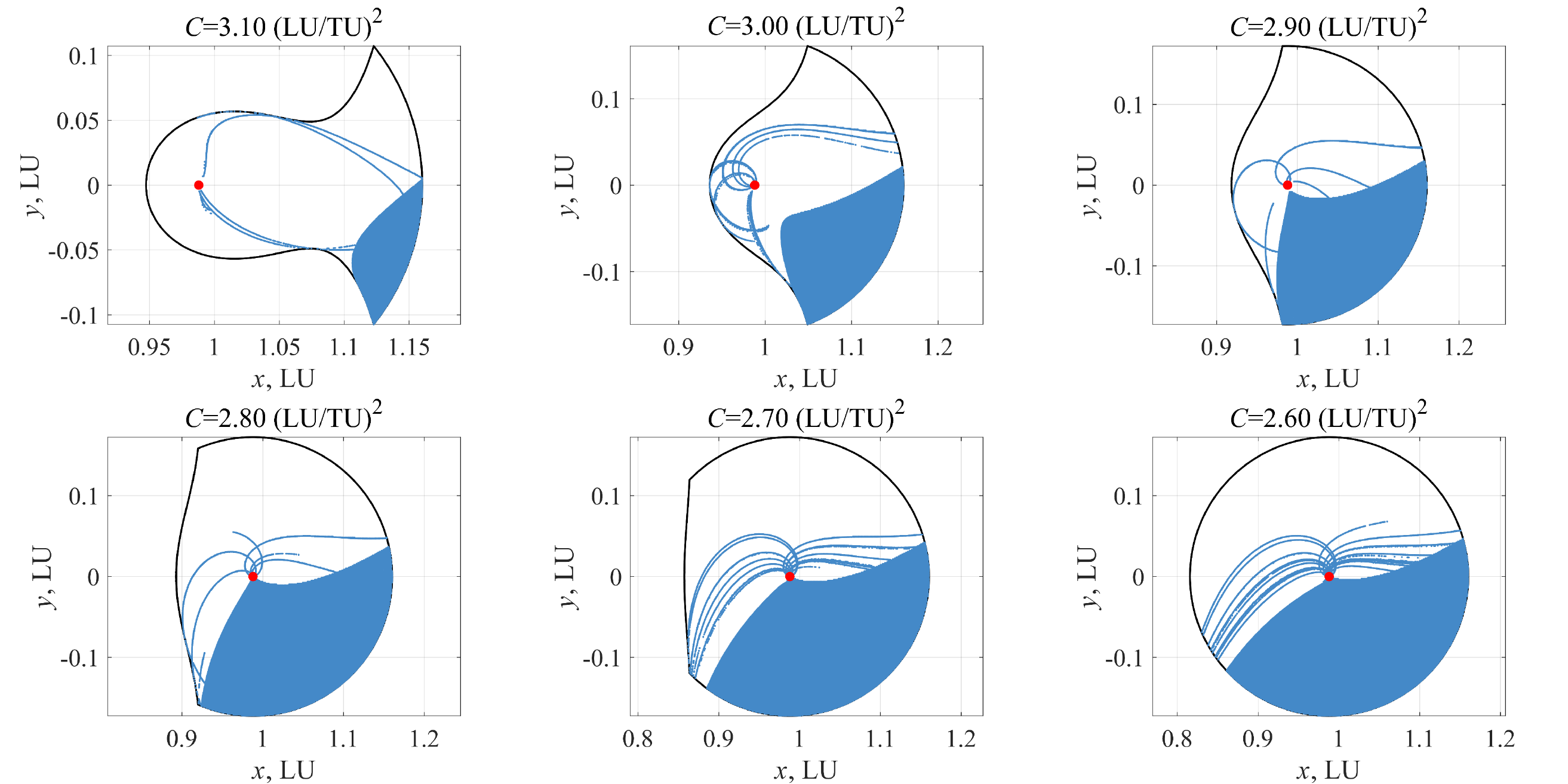}
\caption{Initial states in the ETD satisfying the escape criterion within 100 days.}
\label{initial_states}
\end{figure}

Obtaining the initial states satisfying $E=0$, we perform a forward integration to identify escape trajectories. The maximum integration time is set to 100 days. Escape criterion mentioned in Section \ref{subsec2.3} is adopted to identify escape trajectories. During the integration, the collision trajectories are excluded. After selection, the initial states in the ETD satisfying the escape criterion within 100 days are shown in form of $\left(x,\text{ }y\right)$ in Fig. \ref{initial_states}. The red dot denotes the Moon. 

In Fig. \ref{initial_states}, the black solid line denotes the boundaries of the ETD or SOI of the Moon, and the blue scatters denote the selected initial states. It can be observed that the initial states generating escape trajectories are concentrated in the lower right corner of the ETD, which indirectly confirms the numerical findings of Qi and Xu \cite{qi2015mechanical}. Based on these selected initial states, a backward integration is performed to search the states satisfying Eq. \eqref{eq19}.
\subsubsection{Performing Backward Integration to Search States Satisfying Eq. (19)}\label{subsubsec4.2.3}
\begin{figure}[h]
\centering
\includegraphics[width=0.9\textwidth]{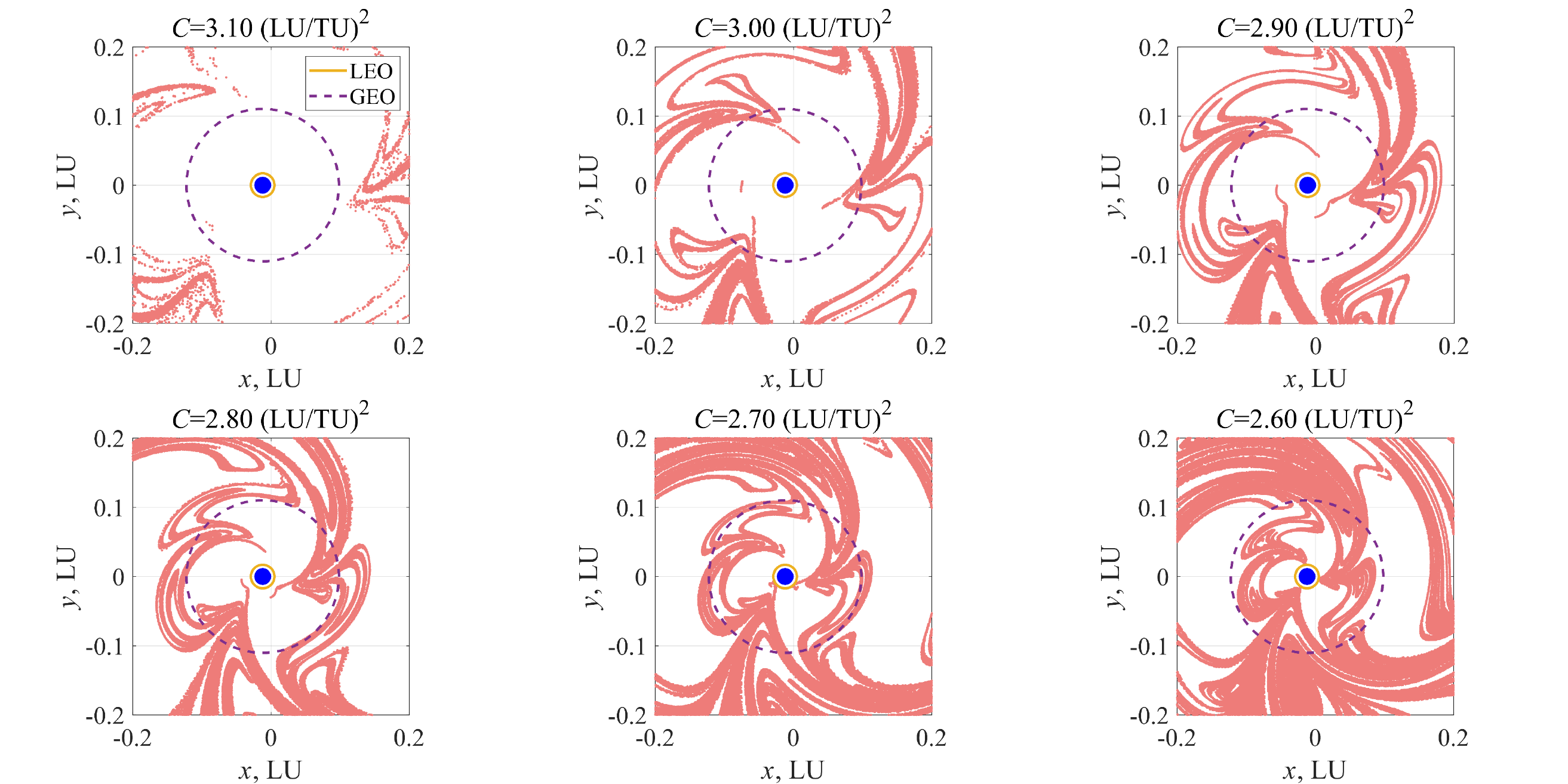}
\caption{Distribution of apsis positions.}
\label{apsis}
\end{figure}

When obtaining initial states generating escape trajectories in Section \ref{subsubsec4.2.2}, we use these initial states to perform a backward integration. The maximum integration time is set to 100 days. During the integration, the collision trajectories are also excluded. The apses with respect to the Earth during this integration are recorded. For each trajectory, a maximum of 20 apses are recorded. The states of apses with respect to the Earth satisfy:
\begin{equation}
{\left( {{x_p} + \mu } \right)\left( {{u_p} - {y_p}} \right) + {y_p}\left( {{v_p} + {x_p} + \mu } \right)}=0 \label{eq26}
\end{equation}
The distribution of apsis positions is presented in Fig. \ref{apsis}. The orange scatters denote the apsis positions, and the blue dot denotes the Earth. When the apsis distribution intersects the Earth parking orbits (LEO and GEO), we consider there exist gravity-assist escape trajectories departing from the corresponding parking orbits. The apsis distribution varies with $C$. Under the investigated range of $C$, when $C\leq 3.0\text{ }{\left( {{\text{LU/TU}}} \right)^2}$, the apsis distribution intersects the GEO. Moreover, when $C\leq 2.7\text{ }{\left( {{\text{LU/TU}}} \right)^2}$, the apsis distribution intersects the LEO. Because $\Delta{v}_i$ directly corresponds to $C$ \cite{campagnola2010endgame}, we only select the case under $C=3.0\text{ }{\left( {{\text{LU/TU}}} \right)^2}$ to construct gravity-assist escape trajectories departing from the GEO and the case under $C=2.7\text{ }{\left( {{\text{LU/TU}}} \right)^2}$ to construct gravity-assist escape trajectories departing from the LEO. We record the apsis position from the Earth satisfying $r_{1p}\in \left[R_\text{E}+h_{ij}- 5,\text{ }R_\text{E}+h_{ij}+ 5\right] \text{ } \left(\text{km}\right)\text{ }\left(j=1,\text{ }2\right)$ and the corresponding $\left(u_p,\text{ }v_p\right)$ as the departure states of gravity-assist escape trajectories (note that although there exist apses satisfying $r_{1p}<R_\text{E}+h_{i2}$ under $C=3.1\text{ }{\left( {{\text{LU/TU}}} \right)^2}$, the corresponding apses do not satisfy $r_{1p}\in \left[R_\text{E}+h_{i2}- 5,\text{ }R_\text{E}+h_{i2}+ 5\right] \text{ } \left(\text{km}\right)$). For each selected escape trajectory, the time interval from the departure epoch to the epoch when the escape criterion are satisfied is defined as the time of flight (TOF). Subsequently, the construction results are presented and analyzed.

\textit{Remark}: Following the aforementioned method, the gravity-assist escape trajectories can be constructed effectively. Compared to grid search method \cite{FU2025}, this method introduces some prior knowledge about the range of $C$ (i.e., when $C<C^*$, the two regions of ETD merge into one region). Moreover, the method in Ref. \cite{FU2025} was designed to search for escape trajectories generally, rather than specifically for gravity-assist escape trajectories, and obtained trajectories also included direct escape trajectories. In contrast, the method proposed in this Note can construct gravity-assist escape trajectories in a targeted way, because the construct parameters of escape trajectories are selected in the SOI of the Moon.

\subsection{Results and Discussion}\label{subsec4.3}
The gravity-assist escape trajectories are constructed by the aforementioned method. Because under the same value of $C$, the value of $\Delta{v}_i$ varies slightly in the given Earth parking orbit \cite{campagnola2010endgame}, we select the solutions with the minimum TOF for the two cases: escape trajectories departing from the LEO and GEO.

\begin{figure}[h]
\centering
\includegraphics[width=0.8\textwidth]{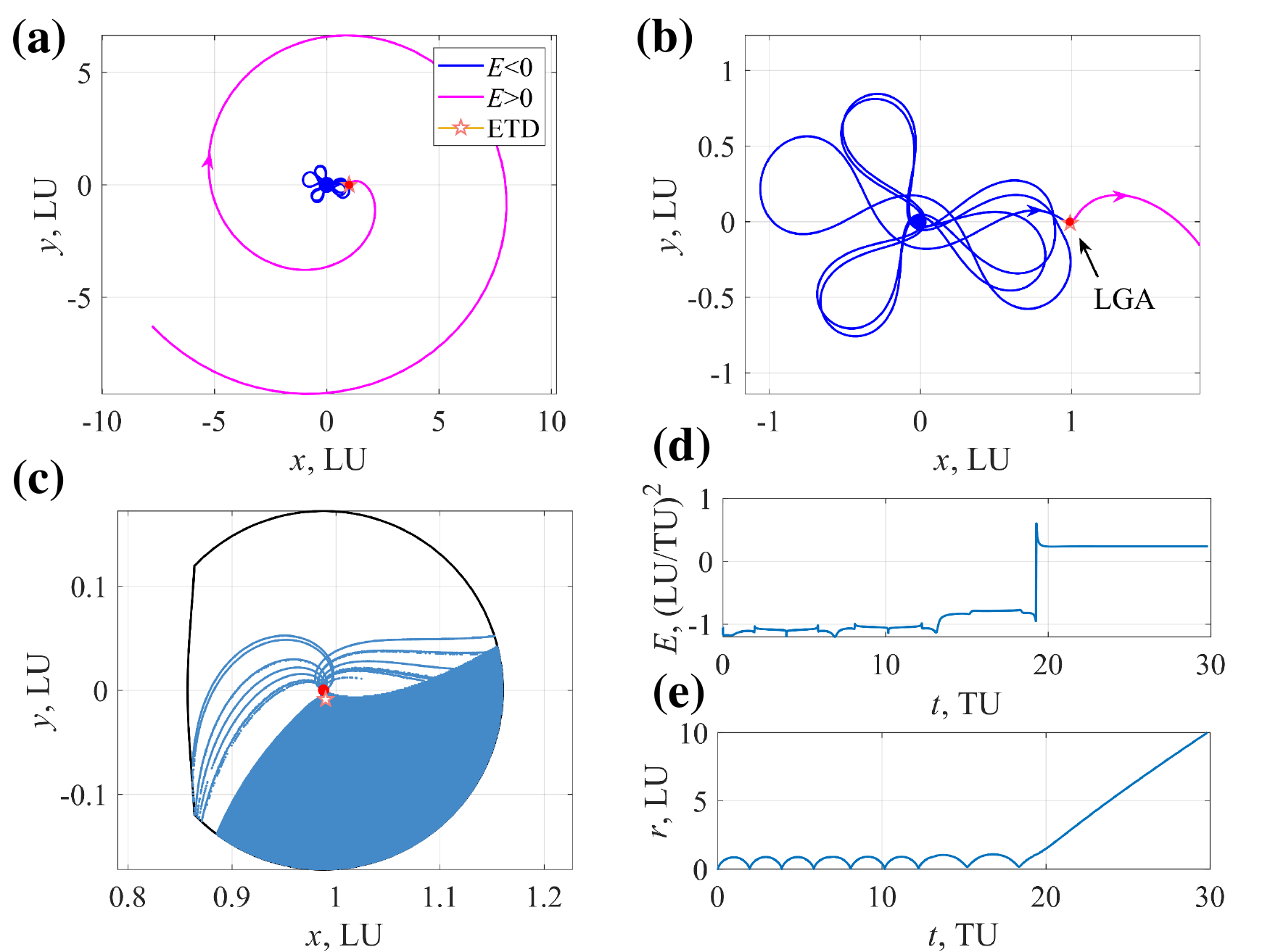}
\caption{The obtained solution departing from the LEO. (a) Trajectory; (b) Detailed trajectory; (c) Corresponding states in the ETD; (d) Variation in $E$; (e) Variation in $r$.}
\label{LEO}
\end{figure}

Figure \ref{LEO} presents the corresponding information about the solution departing from the LEO with the minimum TOF. The orange star denotes the states corresponding to the ETD. In Fig. \ref{LEO} (b), it is observed that the trajectory completes multiple revolutions around the Earth before the LGA. The value of $E$ varies from negative to positive after the LGA. The type of the LGA is a direct LGA \cite{qi2015mechanical}, i.e., during the LGA, the angular momentum with respect to the Moon is positive.

\begin{figure}[h]
\centering
\includegraphics[width=0.8\textwidth]{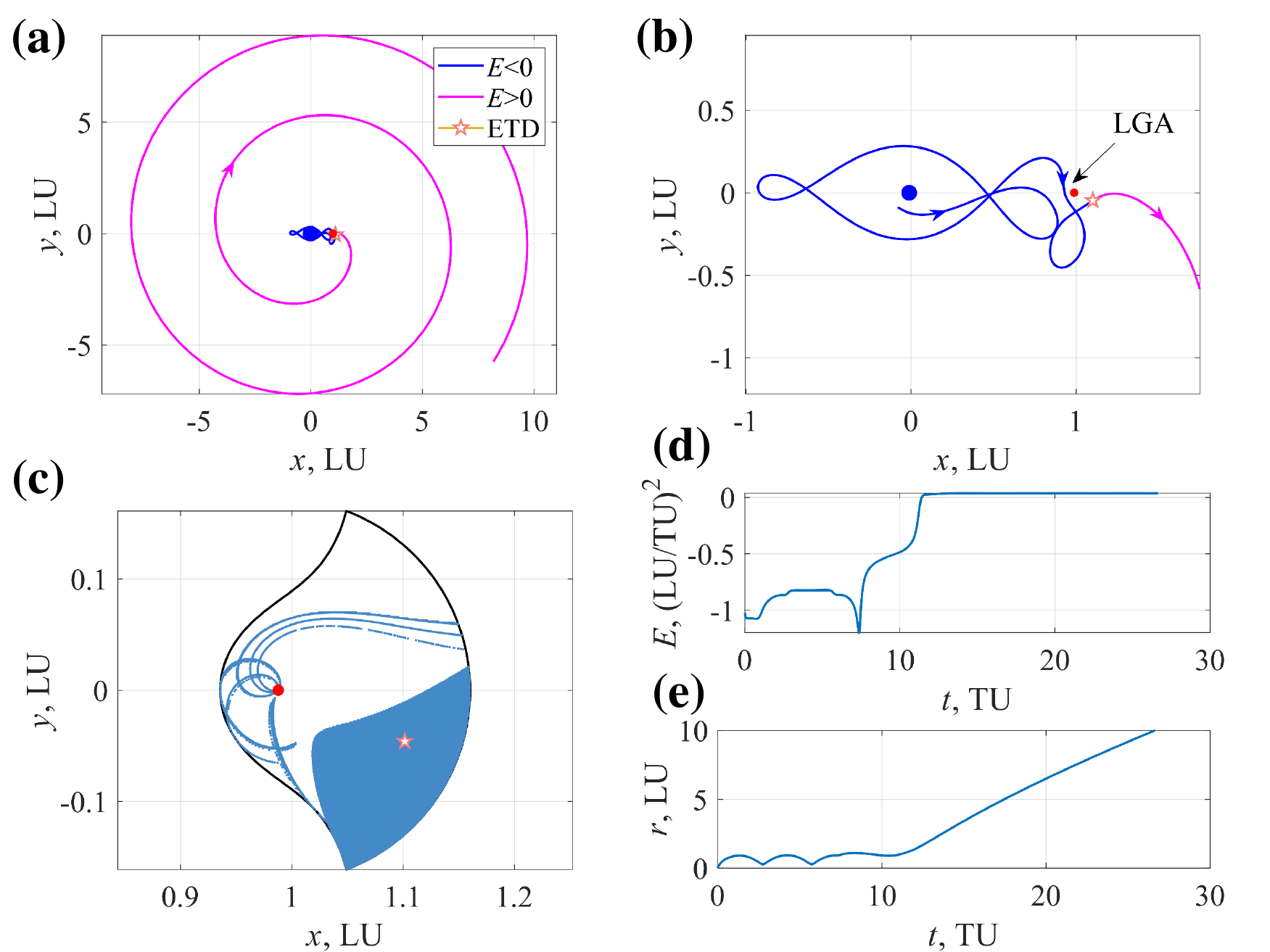}
\caption{The obtained solution departing from the GEO. (a) Trajectory; (b) Detailed trajectory; (c) Corresponding states in the ETD; (d) Variation in $E$; (e) Variation in $r$.}
\label{GEO}
\end{figure}

Figure \ref{GEO} presents the corresponding information about the solution departing from the GEO with the minimum TOF. The trajectory also completes multiple revolutions around the Earth before the LGA. The type of the LGA is also a direct LGA. From these two solutions, it confirms that $E$ varies more significantly during the Moon flyby than during the Earth flyby \cite{qi2015mechanical}. Then, the $\Delta{v}_i$ and TOF of these two solutions are presented in Table \ref{tab2}. Moreover, the minimum $\Delta{v}_i$ of direct escape is estimated based on the Earth-centered two-body model \cite{fu2025four}. Compared to the $\Delta{v}_i$ of direct escape, it confirms that LGA can help in the reduction of the impulse of escape trajectories.

\begin{table}[h]
\centering
\renewcommand{\arraystretch}{1.5}
\caption{Information about the obtained two solutions.}\label{tab2}%
\begin{tabular}{@{}lllll@{}}
\hline
Solution    & $\Delta{v}_i,\text{ }\text{km/s}$  & TOF, Day & Earth Parking Orbit \\
\hline
I & $3.121$  & 130 & LEO \\
II & $1.009$  & 116 &  GEO\\
Direct Escape &$3.228$ & -- & LEO \\
Direct Escape &$1.269$ & -- & GEO \\
\hline
\end{tabular}
\end{table}

\section{Conclusion}\label{sec5}
This Note is devoted to developing the concept and theory of energy transition domain (ETD) defined by the mechanical energy of spacecraft in the Earth-Moon planar circular restricted three-body problem (PCR3BP) and constructing gravity-assist escape trajectories based on it. Firstly, the concept of the ETD is presented, and its dependency on the Jacobi energy is analyzed. It is found that there exists a critical value of the Jacobi energy. When the value of the Jacobi energy is higher than this value, the ETD is mainly divided into two regions: one in the Moon region and the other in the Earth-Moon exterior region. When the value of the Jacobi energy is lower than this value, the two regions merge into one region. This phenomenon may provide prior knowledge about selecting the range of the Jacobi energy in the construction of escape trajectories. Then, an effective method to construct gravity-assist escape trajectories is proposed. Trajectories departing from the 167 km low Earth orbit and 36000 km geosynchronous Earth orbit are considered. The initial states are selected in the ETD in the sphere of influence of the Moon, and the trajectories are searched and generated from the forward and backward integration. Finally, solutions with the minimum time of flight are presented and analyzed. This work extends the theoretical foundation of multi-body escape dynamics, establishing a link between lunar gravity assist and escape trajectories.
\section*{Appendix A: Approximate Relationship between $E$ and Two-Body Energy with respect to The Earth-Moon Barycenter}
According to Ref. \cite{FU2025}, the two-body energy with respect to the Earth-Moon barycenter ($E_2$) is defined as:
\begin{equation}
E_2 = \frac{1}{2}\left( {{x^2} + {y^2} + {u^2} + {v^2}} \right) - uy + xv - \frac{{1 }}{{{r}}} \label{eq27}
\end{equation}
When $r$ is sufficiently large (i.e., $r>10\text{ LU}$), the influence of the Earth and Moon can be approximated as that of the Earth-Moon barycenter with unit dimensionless mass \cite{qi2018short,FU2025}. Therefore, under this condition, $E_2$ can be approximated as a constant. The difference between $E$ and $E_2$ can expressed as \cite{FU2025}:
\begin{align}\label{eq28}
    E - E_2 & =  - \frac{{1 - \mu }}{{{r_1}}} - \frac{\mu }{{{r_2}}} + \frac{1}{r} \\ 
  &\notag =  - \left( {1 - \mu } \right)\left( {\frac{1}{{{r_1}}} - \frac{1}{r}} \right) - \mu \left( {\frac{1}{{{r_2}}} - \frac{1}{r}} \right) \\ 
  &\notag =  - \frac{{1 - \mu }}{r}\left( {\frac{1}{{\sqrt {1 + {\gamma _1}} }} - 1} \right) - \frac{\mu }{r}\left( {\frac{1}{{\sqrt {1 + {\gamma _2}} }} - 1} \right) 
\end{align}
where:
\begin{equation}
{\gamma _1} = \frac{{2\mu x + {\mu ^2}}}{{{r^2}}} \text{ }\text{ }\text{ }{\gamma _2} = \frac{{2(\mu-1) x + {(\mu -1)^2}}}{{{r^2}}}
\label{eq29}
\end{equation}
Using the Taylor expansion with Lagrange remainder, Eq. \eqref{eq28} can be approximated as:
\begin{align}\label{eq30}
    E - E_2 & =  - \frac{{1 - \mu }}{r}\left[ {{{\left( {1 + {\gamma _1}} \right)}^{ - \frac{1}{2}}} - 1} \right] - \frac{\mu }{r}\left[ {{{\left( {1 + {\gamma _2}} \right)}^{ - \frac{1}{2}}} - 1} \right] \\ 
  \notag & =  - \frac{{1 - \mu }}{r}\left( { - \frac{1}{2}{\gamma _1} + \frac{3}{8}{\gamma _1}^2}-\frac{5}{16}\frac{{\gamma _1}^3}{{\left(1+\theta_1\gamma_1\right)}^{7/2}} \right) - \frac{\mu }{r}\left( { - \frac{1}{2}{\gamma _2} + \frac{3}{8}{\gamma _2}^2}-\frac{5}{16}\frac{{\gamma _2}^3}{{\left(1+\theta_2\gamma_2\right)}^{7/2}} \right)
\end{align}
where $\theta_1 \in \left(0,\text{ }1\right)$ and $\theta_2 \in \left(0,\text{ }1\right)$. Equation \eqref{eq30} can be further expressed as:
\begin{align}\label{eq31}
E-E_2 &= \frac{\mu\left(1-\mu\right)}{2r^3}-\frac{3\mu\left(1-\mu\right)x^2}{2r^5}+ \frac{{3\left( {1 - 2\mu } \right)\mu \left( {1 - \mu } \right)x}}{{2{r^5}}} \\
   \notag &- \frac{{3\mu \left( {1 - \mu } \right)\left( {1 - 3\mu  + 3{\mu ^2}} \right)}}{{8{r^5}}} 
+\frac{5}{16}\frac{{1 - \mu }}{r}\frac{{\gamma _1}^3}{{\left(1+\theta_1\gamma_1\right)}^{7/2}}+\frac{5}{16}\frac{{ \mu }}{r}\frac{{\gamma _2}^3}{{\left(1+\theta_2\gamma_2\right)}^{7/2}} \\
\notag &= \frac{{\mu \left( {1 - \mu } \right){y^2} - 2\mu \left( {1 - \mu } \right){x^2}}}{{2{r^5}}} + \frac{{3\left( {1 - 2\mu } \right)\mu \left( {1 - \mu } \right)x}}{{2{r^5}}} \\
   \notag &- \frac{{3\mu \left( {1 - \mu } \right)\left( {1 - 3\mu  + 3{\mu ^2}} \right)}}{{8{r^5}}} 
+\frac{5}{16}\frac{{1 - \mu }}{r}\frac{{\gamma _1}^3}{{\left(1+\theta_1\gamma_1\right)}^{7/2}}+\frac{5}{16}\frac{{ \mu }}{r}\frac{{\gamma _2}^3}{{\left(1+\theta_2\gamma_2\right)}^{7/2}}
\end{align}
Then, we can obtain:
\begin{align}\label{eq32}
   \left| {E - {E_2}} \right| &\leq \frac{{3\mu }}{{2{r^3}}} + \frac{{3\mu \left| x \right|}}{{2{r^5}}} + \frac{{3\mu }}{{8{r^5}}} + \left| {  \frac{5}{{16}}\frac{{1 - \mu }}{r}\frac{{{\gamma _1}^3}}{{{{\left( {1 + {\theta _1}{\gamma _1}} \right)}^{7/2}}}} + \frac{5}{{16}}\frac{\mu }{r}\frac{{{\gamma _2}^3}}{{{{\left( {1 + {\theta _2}{\gamma _2}} \right)}^{7/2}}}}} \right| \hfill \\
   \notag&\leq \frac{{3\mu }}{{2{r^3}}} + \frac{{3\mu }}{{2{r^4}}} + \frac{{3\mu }}{{8{r^5}}} + \left| {  \frac{5}{{16}}\frac{{1 - \mu }}{r}\frac{{{\gamma _1}^3}}{{{{\left( {1 + {\theta _1}{\gamma _1}} \right)}^{7/2}}}} + \frac{5}{{16}}\frac{\mu }{r}\frac{{{\gamma _2}^3}}{{{{\left( {1 + {\theta _2}{\gamma _2}} \right)}^{7/2}}}}} \right|
\end{align}
Then, we consider the inequalities $\gamma_1$ and $\gamma_2$ satisfy:
\begin{equation}
\left| {{\gamma _1}} \right| \leqslant \frac{{2\mu \left| x \right|}}{{{r^2}}} + \frac{{{\mu ^2}}}{{{r^2}}} \leqslant \frac{{2\mu }}{r} + \frac{{{\mu ^2}}}{{{r^2}}}
\label{eq33}
\end{equation}
\begin{equation}
\left| {{\gamma _2}} \right| \leq \frac{{2\left( {1 - \mu } \right)\left| x \right|}}{{{r^2}}} + \frac{{{{\left( {1 - \mu } \right)}^2}}}{{{r^2}}} \leq \frac{2}{r} + \frac{1}{{{r^2}}}
\label{eq34}
\end{equation}
When $r>10\text{ LU}$, we can obtain:
\begin{equation}
\left| {{\gamma _1}} \right| \leq 2.01\frac{\mu }{r} \leq 0.002613\text{ }\text{ }\text{ }\left| {{\gamma _2}} \right| \leq \frac{{2.1}}{r} \leq 0.21
\label{eq35}
\end{equation}
Therefore:
\begin{equation}
\left| {\frac{5}{{16}}\frac{{1 - \mu }}{r}\frac{{{\gamma _1}^3}}{{{{\left( {1 + {\theta _1}{\gamma _1}} \right)}^{7/2}}}} + \frac{5}{{16}}\frac{\mu }{r}\frac{{{\gamma _2}^3}}{{{{\left( {1 + {\theta _2}{\gamma _2}} \right)}^{7/2}}}}} \right| \leq \frac{{7\mu }}{{{r^4}}}
\label{eq36}
\end{equation}
Then, we can obtain the following inequality: 
\begin{equation}
\left| {E - {E_2}} \right| \leq \frac{{3\mu }}{{2{r^3}}} + \frac{{3\mu }}{{2{r^4}}} + \frac{{3\mu }}{{8{r^5}}} + \frac{{7\mu }}{{{r^4}}} \leq \frac{{4\mu }}{{{r^3}}}
\label{eq37}
\end{equation}
Here, the approximate relationship between $E$ and $E_2$ is derived. This relationship was perviously explored in Ref. \cite{FU2025}. However, the derivation of this relationship in Ref. \cite{FU2025} contains some minor inaccuracies, and the accurate form of the relationship when $r>10\text{ LU}$ is presented in this appendix. According to this derivation, it can be found that the inequality $\left| {E - {E_2}} \right| \leq \frac{{4\mu }}{{{r^3}}}$ obtained in Ref. \cite{FU2025} still hold when $r>10\text{ LU}$.

\section*{Appendix B: An Example of Escape Trajectories under $C>C^*$}
\begin{figure}[h]
\centering
\includegraphics[width=0.8\textwidth]{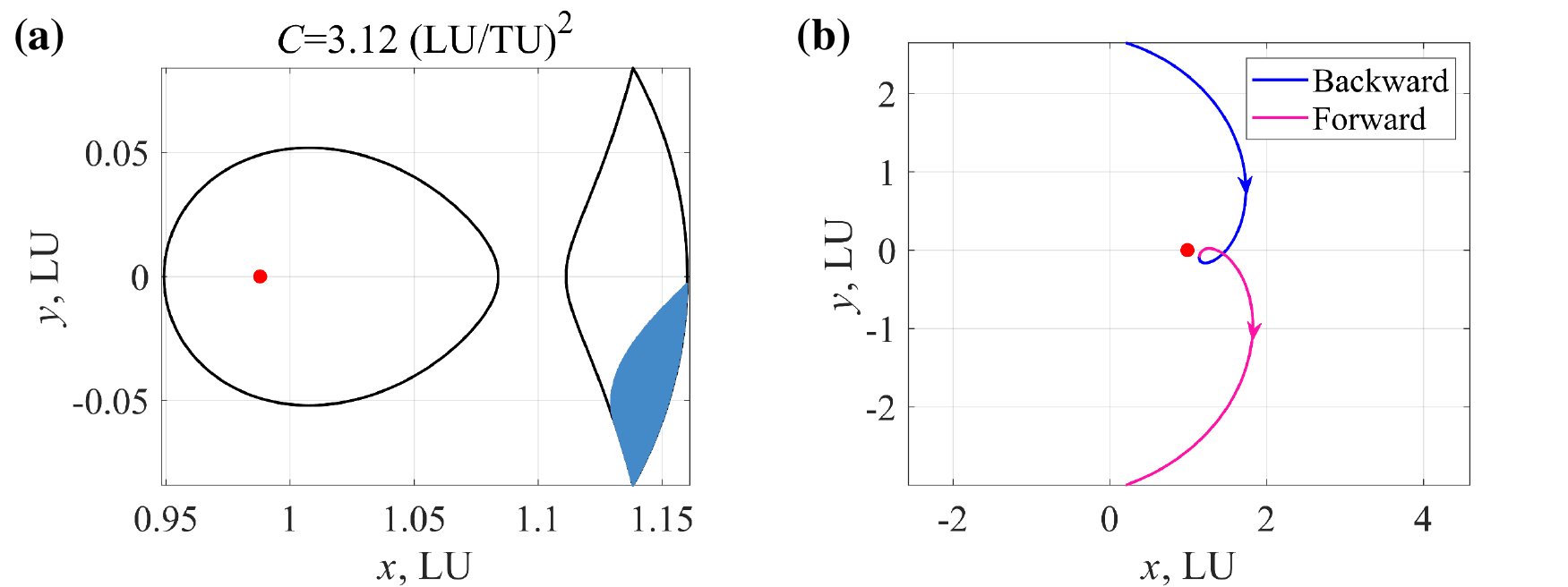}
\caption{Escape trajectory under $C>C^*$. (a) Initial states in the ETD; (b) A sample of escape trajectories.}
\label{Not_effective}
\end{figure}

In this appendix, we select the value of $C$ as $C=3.12\text{ } {\left( {{\text{LU/TU}}} \right)^2}$ (here $C>C^*$), and generate the initial states in the corresponding ETD by the method mentioned in Section \ref{subsubsec4.2.1}. Then, we perform a forward integration of these initial states within 100 days, and the initial states generating escape trajectories are labeled as blue scatters in Fig. \ref{Not_effective} (a). It can be observed that states in the ETD in the Moon region can not generate escape trajectories. By performing a backward integration of the initial states generating escape trajectories, we find that all the trajectories are similar to the sample shown in Fig. \ref{Not_effective} (b). These trajectories remain primarily in the Earth-Moon exterior region and therefore can not be considered as typical gravity-assist escape trajectories departing from Earth parking orbits. Notably, this example only confirms the implication that there may not be trajectories escaping from the Earth region effectively, whether counterexamples exist remains an open question.

\section*{Funding Sources}
The authors acknowledge the financial support from the National Natural Science Foundation of China (Grant No. 12372044), the National Natural Science Foundation of China (No. U23B6002), the National Natural Science Foundation of China (Grant No. 12302058), and the Postdoctoral Science Foundation of China (Grant No. 2024T170480).

\end{document}